\documentclass[sigconf]{acmart}
\usepackage{balance} 
\AtBeginDocument{%
  }

\copyrightyear{2026}
\acmYear{2026}
\setcopyright{cc}
\setcctype{by}
\acmConference[SIGIR '26] {Proceedings of the 49th International ACM SIGIR Conference on Research and Development in Information Retrieval}{July 20--24, 2026}{Melbourne, VIC, Australia.}
\acmBooktitle{Proceedings of the 49th International ACM SIGIR Conference on Research and Development in Information Retrieval (SIGIR '26), July 20--24, 2026, Melbourne, VIC, Australia}
\acmISBN{979-8-4007-2599-9/2026/07}
\acmDOI{10.1145/3805712.3808481}

\begin{document}

\title{Joint Optimization of Relevance and Engagement in Multi-Task Ranking for E-Commerce with Efficient LLM Supervision}


\author{Luming Chen}
\affiliation{%
  \institution{DoorDash Inc.}
  \city{San Francisco}
  \state{California}
  \country{USA}
}
\email{luming.chen@doordash.com}

\author{Jiaqi Xi}
\affiliation{%
  \institution{DoorDash Inc.}
  \city{San Francisco}
  \state{California}
  \country{USA}
}
\email{jiaqi.xi@doordash.com}

\author{Raghav Saboo}
\affiliation{%
  \institution{DoorDash Inc.}
  \city{San Francisco}
  \state{California}
  \country{USA}
}
\email{raghav.saboo@doordash.com}

\author{Kenny Chi}
\affiliation{%
  \institution{DoorDash Inc.}
  \city{San Francisco}
  \state{California}
  \country{USA}
}
\email{kenny.chi@doordash.com}

\author{Martin Wang}
\affiliation{%
  \institution{DoorDash Inc.}
  \city{San Francisco}
  \state{California}
  \country{USA}
}
\email{martin.wang@doordash.com}

\author{Sudeep Das}
\affiliation{%
  \institution{DoorDash Inc.}
  \city{San Francisco}
  \state{California}
  \country{USA}
}
\email{sudeep.das2@doordash.com}

\author{Danny Nightingale}
\affiliation{%
  \institution{DoorDash Inc.}
  \city{San Francisco}
  \state{California}
  \country{USA}
}
\email{danny.nightingale@doordash.com}

\author{Aditya Dodda}
\affiliation{%
  \institution{DoorDash Inc.}
  \city{San Francisco}
  \state{California}
  \country{USA}
}
\email{aditya.dodda@doordash.com}

\author{Elyse Winer}
\affiliation{%
  \institution{DoorDash Inc.}
  \city{San Francisco}
  \state{California}
  \country{USA}
}
\email{elyse.winer@doordash.com}

\author{Akshad Viswanathan}
\affiliation{%
  \institution{DoorDash Inc.}
  \city{San Francisco}
  \state{California}
  \country{USA}
}
\email{akshad.viswanathan@doordash.com}

\renewcommand{\shortauthors}{Chen et al.}

\begin{abstract}
Optimizing industrial search ranking models solely for user engagement signals often introduces systematic biases, prioritizing popular or price-anchored items that may not satisfy semantic intent. We present a production-scale multi-task ranking system that integrates semantic relevance as a primary optimization objective, enabling explicit and controllable relevance--engagement trade-offs. Our architecture employs an ordinal relevance head that predicts cumulative probabilities over relevance thresholds, preserving the inherent ordering of labels. These outputs are integrated with engagement heads through a unified value model scoring function, enabling systematic balancing of semantic quality and short-term behavioral signals. To provide high-quality supervision for this multi-task framework, we utilize fine-tuned lightweight Large Language Models (LLMs) to generate three-level ordinal relevance labels: irrelevant, moderately relevant, and highly relevant. We address challenges regarding label distribution sensitivity and ensure high alignment with human annotations to enable efficient labeling for over 100 million query--item pairs. Evaluation across offline metrics, including NDCG@10, and online A/B experiments demonstrates that our approach significantly improves semantic alignment while preserving core engagement objectives.
\end{abstract}





\maketitle

\section{Introduction}
Search is a primary driver of discovery and revenue in large-scale e-commerce platforms. In production systems, ranking models are typically optimized against user engagement signals such as click-through rate (CTR), add-to-cart rate (ATCR), and conversion rate (CVR). These signals are abundant and directly aligned with business objectives \cite{fu2024residual, morishetti2025personalized}. However, optimizing solely for engagement can introduce systematic biases: items that attract interaction due to popularity, price anchoring, or lexical overlap may be up-ranked even when they do not semantically satisfy user intent \cite{hager2024unbiased,joachims2017unbiased}. Over time, such behavior can degrade perceived search quality, increase query reformulation, and erode user trust.

A central challenge in addressing this issue is the scarcity of high-quality semantic relevance supervision at production scale. Human annotation is expensive, slow, and difficult to maintain for dynamic catalogs containing millions of items and long-tail queries. Behavioral signals are abundant but are affected by position bias, exposure bias, and feedback loops \cite{hager2024unbiased,joachims2017unbiased}. As a result, many industrial ranking systems lack reliable fine-grained semantic supervision beyond implicit engagement feedback.

Recent advances in large language models (LLMs) offer a promising alternative. Multiple studies show that prompted or fine-tuned LLMs can generate high-quality relevance judgments that closely align with human annotations in product search and web-scale settings \cite{liu2024_towards_more_relevant_product_search,mehrdad2024llm_relevance_product_search,patil2025llm_pinterest,zhuang2024beyond,wang2024improving}. However, integrating LLM supervision into production search systems presents practical challenges. First, LLM inference is too expensive and latency-sensitive for real-time serving at high queries-per-second rates. Second, fine-tuned LLM classifiers can be sensitive to label distribution and may systematically over- or under-predict certain relevance levels. Third, raw LLM scores must be calibrated and structured appropriately to interact with existing engagement-driven ranking objectives \cite{rahmani2025judging}.

In this work, we present a deployed multi-task ranking system that incorporates efficient LLM-based relevance supervision as a first-class optimization objective. We fine-tune GPT models on hundreds of thousands of human-annotated query--item pairs and generate large-scale three-level ordinal relevance labels offline. We observe that fine-tuning performance is highly sensitive to the underlying label distribution, and we adopt sampling strategies that mimic production relevance proportions to ensure stable classification behavior. The resulting model achieves strong agreement with human annotations, including on long-tail and multi-word queries.

We integrate the LLM-generated relevance signal into a shared multi-task neural ranking architecture that jointly predicts CTR, ATCR, CVR, and relevance. A unified value function enables explicit control over relevance and engagement trade-offs at serving time. This design allows us to improve semantic alignment while maintaining core business performance, without invoking LLMs during online inference.

Our contributions are threefold:
\begin{itemize}
\item We design a scalable pipeline for LLM-based relevance supervision that generates high-quality ordinal labels aligned with production query distributions.
\item We integrate ordinal semantic relevance directly into a multi-task ranking objective, enabling controllable trade-offs between relevance and engagement.
\item We demonstrate offline relevance improvements and positive online impact in a large-scale production search system.
\end{itemize}

\section{Related Work}
Multi-task learning (MTL) is widely adopted in industrial product search ranking to jointly optimize engagement-driven objectives such as CTR, ATCR, and CVR using shared representations with task-specific heads. Recent systems integrate heterogeneous signals, including tabular user features and semantic text embeddings, to model complex query--item interactions, consistently outperforming single-task and tree-based approaches \cite{wu2022multi,fu2024residual,morishetti2025personalized}. However, relevance is often absent from the primary optimization objectives or treated as an auxiliary task, leaving such models susceptible to bias and exposure effects.

LLMs have recently demonstrated strong capability in generating high-quality relevance supervision at scale. Prior work demonstrates that fine-tuned LLMs can produce relevance judgments with strong agreement with human annotations, enabling large-scale offline evaluation and label generation for training and distillation \cite{mehrdad2024llm_relevance_product_search, zhuang2024beyond, patil2025llm_pinterest, liu2024_towards_more_relevant_product_search, wang2024improving,agrawal2025rationale}. LLM-predicted semantic relevance scores have been incorporated into learning-to-rank objectives to complement engagement-based signals through calibrated label formulations that balance content relevance and behavioral outcomes \cite{liu2024_towards_more_relevant_product_search}. In addition, several studies highlight the importance of fine-grained relevance supervision, showing that graded relevance labels improve ranking reliability compared to binary judgments \cite{zhuang2024beyond, patil2025llm_pinterest, wang2024improving}. Separately, LLMs have also been explored as reranking agents, showing strong effectiveness but high inference cost \cite{sun2023chatgpt,zhuang2024setwise}.

In contrast to prior work that treats LLMs primarily as external relevance teachers, evaluation tools, or standalone rerankers, our approach integrates efficient, fine-grained LLM supervision directly into a multi-task ranking objective, enabling explicit control of relevance--engagement trade-offs in production-scale search systems without incurring online LLM inference overhead.

\section{Methodology}
Given a user query and a set of candidate items, our goal is to rank items by jointly optimizing relevance and user engagement outcomes. Engagement is modeled through multiple behavioral signals, including CTR, ATCR, and CVR, while relevance reflects the semantic match between query intent and item content. 
\subsection{Multi-task Model Architecture}
\begin{figure}[htbp]
  \includegraphics[width=0.9\columnwidth]{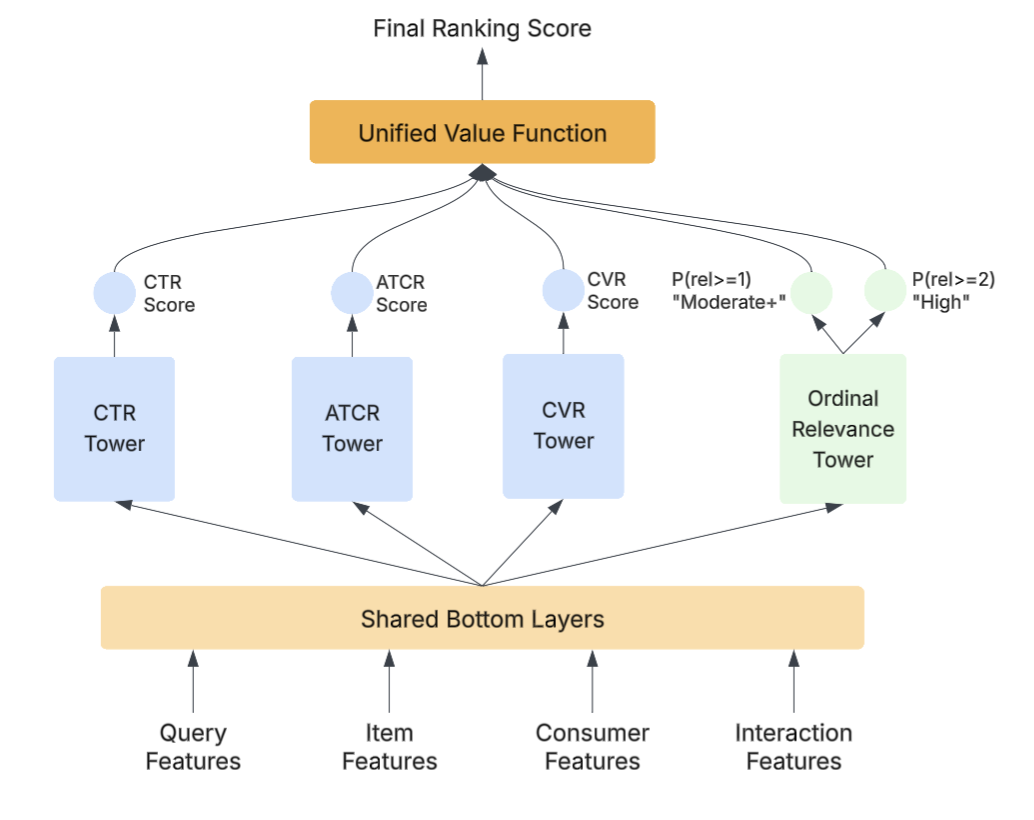}
  \caption{Multi-task ranking architecture with unified relevance--engagement optimization}
  \Description{}
  \label{fig:multitask model}
\end{figure}

We adopt a multi-task deep neural network (DNN) architecture that jointly predicts engagement and relevance signals, as illustrated in Figure \ref{fig:multitask model}. The model incorporates multiple feature groups, including query features (e.g., query embeddings, inferred categories, and brand), item features (e.g., item embeddings, price, category, brand attributes, and historical engagement statistics), and consumer features representing past user interactions. In addition, we incorporate interaction features between queries and items, such as BM25 scores for multiple text fields, historical query--item engagement rates, and string similarity measures between the query and item text attributes. We also include user--item and user--taxonomy interaction signals derived from historical engagement patterns to capture personalized relevance. These features are processed by shared bottom layers and then fed into task-specific towers for engagement and relevance prediction.

The relevance score can serve multiple purposes: it can be used for result filtering or categorical separation in the UI, and it acts as a critical component to improve the semantic quality of the final ranking. Integrating this relevance head introduces negligible additional computational cost due to the following structural efficiencies:

\begin{itemize}
    \item \textbf{Shared Bottom Layers}: The relevance head shares the bottom layers with the engagement tasks, where the primary computational costs--specifically high-cardinality embedding lookups--are generated.
    \item \textbf{Feature Reusability}: The relevance task utilizes a subset of the features already being processed for the engagement models, avoiding duplicated feature fetching.
    \item \textbf{Single-Pass Inference}: By consolidating these objectives into one model, we avoid the need for separate service calls or multiple forward passes.
\end{itemize}

\subsection{Fine-grained Relevance Modeling}
Let $r_{q,i}$ denote the semantic relevance level between a query $q$ and an item $i$. We adopt a three-level ordinal relevance formulation,
$r_{q,i} \in \{0, 1, 2\}$, where 0 denotes irrelevant, 1 denotes moderately relevant, and 2 denotes highly relevant. Compared to binary relevance judgments, this graded formulation strikes a balance between capturing nuanced query and item relationships and maintaining labeling reliability. In the context of e-commerce search, distinguishing irrelevant items from relevant ones is essential for user trust, while identifying highly relevant results is critical for consistently prioritizing the best matches under competing business objectives.

\subsubsection{Ordinal Relevance Head}
To reflect the ordinal relevance structure in the model, the relevance head predicts two probabilities, $p(r_{q,i} \ge 1)$ and $p(r_{q,i} \ge 2)$, corresponding to the probability that an item is at least relevant and highly relevant, respectively. This ordinal formulation preserves the natural ordering of relevance levels and allows the model to more effectively separate irrelevant, moderately relevant, and highly relevant items, leading to improved relevance-aware ranking behavior. In addition, the probabilistic outputs are well-suited for calibrated decision thresholds for relevance filtering and quality-based item grouping in downstream ranking and presentation layers.

\subsubsection{LLM-based Relevance Supervision}
Following~\cite{xi2026mine}, we generate training labels through a multi-stage refinement process that ensures both quality and efficiency at production scale. First, we collect approximately 600k human annotations for query--item pairs as ground truth. We then audit these human annotations by comparing them directly against user engagement signals. Specifically, for query--item pairs where the human label is 0 (irrelevant) but the ATCR and CVR ranks are in the top 50\% for that query, we re-evaluate using a larger LLM (e.g., GPT-4o \cite{openai2024gpt4o}, o3 \cite{openai2025o3}) with more granular relevance prompts. We further reconcile conflicts between human and audit LLM labels using our Query-to-Taxonomy (Q2T) transformer model trained to predict relevant item categories for each query. If a category appears in Q2T's predicted set, we adopt the higher of the human or audit LLM score, otherwise, we retain the lower. We then fine-tune a gpt-4o-mini model~\cite{openai2024gpt4omini} on the audited human annotations, and apply this lightweight model to generate relevance labels at scale for over 100M query--item pairs. The fine-tuned LLM achieves a three-class accuracy of $89\%$ and a within-1 accuracy (i.e., $\left|\text{human label} - \text{LLM inference}\right| \leq 1$) of $98\%$ compared to human annotations on a hold-out set of 105k query--item pairs. 

\subsection{Relevance-aware Multi-objective Value Function}
The final ranking score for a query--item pair $(q,i)$ is computed by aggregating predicted engagement outcomes and semantic relevance into a unified value function:
\begin{equation}
S(q,i) = \alpha \hat{y}_{\text{CTR}} + \beta \hat{y}_{\text{ATCR}} + \gamma \hat{y}_{\text{CVR}} + (1-\alpha-\beta-\gamma) \hat{s}_{\text{rel}},
\end{equation}
where $\alpha, \beta, \gamma \in [0,1]$ are hyperparameters controlling the relative importance of engagement objectives, and $\hat{y}$ denotes the predicted probability for each behavioral task. The term $\hat{s}_{\text{rel}}$ represents the predicted scalar relevance score derived from the ordinal relevance head:
\begin{equation}
\hat{s}_{\text{rel}}(q,i) = \hat{p}(r_{q,i} \ge 1) + \hat{p}(r_{q,i} \ge 2),
\end{equation}
where $\hat{p}(r_{q,i} \ge k)$ is the probability output of the ordinal head for each relevance threshold $k \in \{1, 2\}$. This formulation ensures that $\hat{s}_{\text{rel}}$ corresponds to the expected relevance grade:
\begin{equation}
\mathbb{E}[r_{q,i}] = \sum_{k=1}^{2} p(r_{q,i} \ge k) = p(r_{q,i} \ge 1) + p(r_{q,i} \ge 2).
\end{equation}
This value-based approach enables explicit control over relevance and engagement trade-offs via the hyperparameters $\alpha, \beta, \text{and } \gamma$, while producing a single continuous ranking score optimized within the multi-task learning framework. In order to tune these parameters, we conducted various offline tests and a value model tuning effort online through multiple treatments.

\section{Experiments and Results}
We evaluate our proposed framework on a large-scale product search dataset to answer the following research questions:
\begin{itemize}
    \item \textbf{RQ1:} Does integrating fine-grained LLM supervision into an MTL framework improve ranking relevance without compromising engagement?
    \item \textbf{RQ2:} How does the ordinal relevance formulation compare to standard classification and regression baselines?
    \item \textbf{RQ3:} Can we effectively calibrate the trade-off between semantic relevance and behavioral engagement?
\end{itemize}

\subsection{Experimental Setup}
We utilize a dataset consisting of 2 million query--item pairs. Relevance labels are generated using a fine-tuned LLM as described in Section 3.2.2. We report AUC and NDCG@10 to measure engagement prediction and ranking quality. For semantic relevance, we compute NDCG@10 based on relevance head predictions.

\begin{table*}[h]
\centering
\caption{Comparison of different modeling approaches on engagement and relevance tasks.}
\label{tab:offline}
\begin{tabular}{lcccccccc}
\hline
 & \multicolumn{3}{c}{\textbf{AUC}} & \multicolumn{4}{c}{\textbf{NDCG@10}} \\ \cmidrule(lr){2-4} \cmidrule(lr){5-8}
\textbf{Model} & \textbf{Click} & \textbf{Add-to-cart} & \textbf{Conversion} & \textbf{Click} & \textbf{Add-to-cart} & \textbf{Conversion} & \textbf{Relevance} \\ \hline
MTL-Engagement & 0.840 & 0.852 & 0.855 & 0.631 & 0.626 & 0.628 & 0.812 \\
MTL-Softmax & 0.838 & 0.851  & 0.854 & 0.628 & 0.624  & 0.625 & 0.961 \\
MTL-Regression & 0.839  & 0.851 & 0.854  & 0.629 & 0.625 & 0.627 & 0.959 \\
Ours (Auxiliary Only) & \multicolumn{6}{c}{\textit{same as below}} & 0.823 \\
\textbf{Ours (Ordinal)} & 0.838 & 0.851 & 0.855 & 0.630 & 0.625 & 0.627 & 0.962 \\
\hline
\end{tabular}
\end{table*}

\begin{table}[htbp]
    \centering
    \caption{Online experiment results}
    \label{tab:online-ab}
    \begin{tabular}{lcc}
        \toprule
        Metric & Relative lift (p-value) \\
        \midrule
        ATCR & $1.16\%$ $(0.01)$ \\
        CVR & $1.10\%$ $(0.01)$\\
        Gross Order Value (GOV) & $0.50\%$ $(0.03)$\\     
        \bottomrule
    \end{tabular}
\end{table}

We compare our model against:
\begin{itemize}
    \item \textbf{MTL-Engagement}: A standard multi-task model optimized only for CTR, ATCR, and CVR.
    \item \textbf{MTL-Softmax}: A multi-task model using a three-way softmax head for relevance classification.
    \item \textbf{MTL-Regression}: A multi-task model treating relevance grades as continuous values optimized via mean squared error (MSE) loss.
    \item \textbf{Ours (Auxiliary Only)}: A multi-task model using our proposed ordinal relevance head as an auxiliary task, without directly using the relevance score at inference time.
\end{itemize}

\subsection{Main Results (RQ1 \& RQ2)}
Table \ref{tab:offline} summarizes engagement prediction performance and the quality of relevance prediction across all model variants. All models with explicit relevance supervision achieve comparable AUC and NDCG on Click, Add-to-cart, and Conversion relative to the engagement-only baseline, demonstrating that relevance optimization can be integrated into multi-task ranking without compromising core behavioral objectives. 

Models with relevance head achieve substantially higher relevance NDCG compared to the engagement-only baseline, with the proposed ordinal formulation performing best. Since the engagement-only model does not include an explicit relevance head, we use its CTR prediction as a proxy relevance score for offline evaluation. These results demonstrate that directly modeling semantic alignment is significantly more effective than relying on engagement signals as implicit relevance proxies.

The Auxiliary Only variant yields only a marginal improvement in relevance NDCG over the engagement-only baseline, indicating that while the relevance task may slightly regularize shared representations, its impact on relevance ranking remains limited when not explicitly incorporated into the final scoring function. This confirms that directly integrating relevance into the unified value model is necessary to realize substantial relevance gains.

\begin{figure}[htbp]
  \includegraphics[width=0.85\columnwidth]{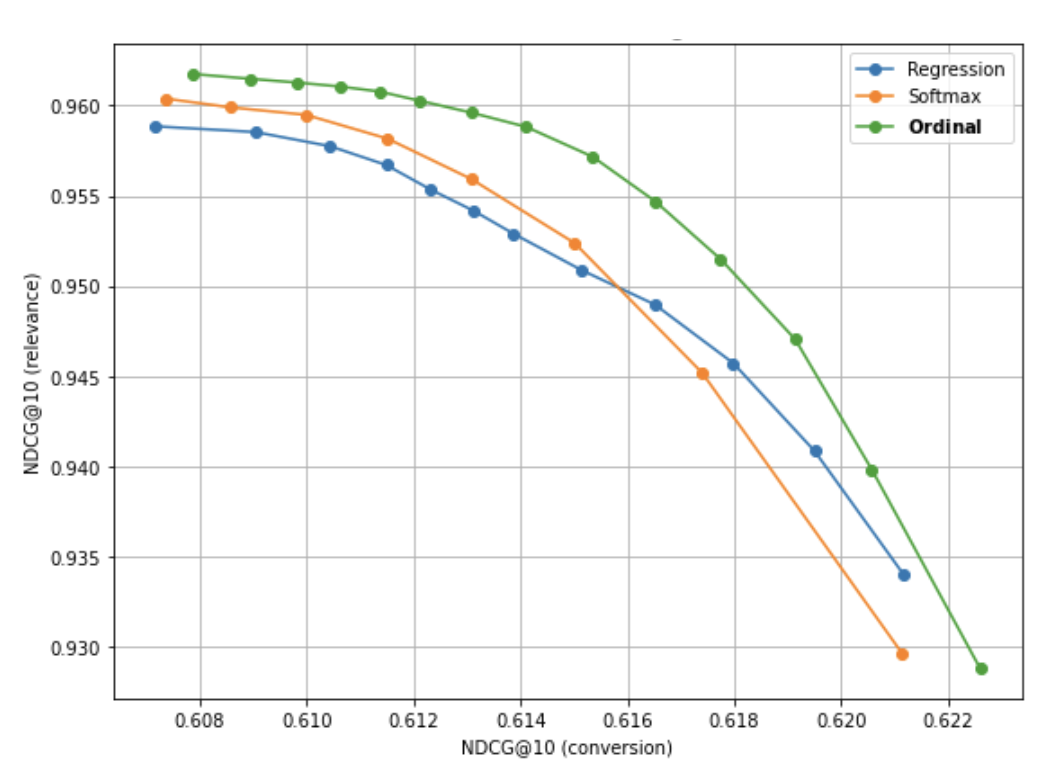}
  \caption{Relevance vs. conversion trade-off (NDCG@10) for different model variants.}
  \Description{}
  \label{fig:tradeoff}
\end{figure}

\subsection{Trade-off Analysis (RQ3)}
\noindent To evaluate the controllability of the proposed relevance-aware ranking formulation, we construct a weighted scoring function that linearly combines the predicted relevance and conversion outputs, and vary the relevance weight. For each weight setting, we compute both relevance NDCG@10 and conversion NDCG@10 on the resulting ranked lists. Figure \ref{fig:tradeoff} plots the corresponding trade-off curves for different model variants. The proposed ordinal formulation consistently yields a more favorable trade-off frontier, achieving higher relevance NDCG at the same level of conversion NDCG compared to softmax and regression baselines. This indicates that the ordinal relevance head produces semantic signals that are more compatible with engagement optimization when integrated through a unified scoring function.

In practice, we observe that modest relevance weights yield substantial relevance gains with minimal engagement loss, enabling practical tuning of semantic quality without sacrificing core engagement objectives.

\subsection{Online A/B Experiment}
We validate our approach through production A/B experiments on live search traffic. The treatment replaces the baseline ranking model with the proposed relevance-aware formulation.

As shown in Table \ref{tab:online-ab}, the relevance-aware model delivers consistent improvements in key business metrics, including add-to-cart rate (ATCR), conversion rate (CVR), and Gross Order Value (GOV). Results are statistically significant at $p < 0.05$ based on an experiment run over three weeks. These results demonstrate that directly incorporating ordinal semantic relevance into the unified value function not only improves offline relevance metrics but also translates into measurable online gains.

\section{Conclusion}
We presented a production-scale multi-task ranking framework that integrates fine-grained LLM-based semantic relevance supervision as a first-class optimization objective. By generating large-scale ordinal relevance labels offline and embedding an ordinal relevance head within a unified value-based ranking formulation, our approach enables explicit and controllable relevance--engagement trade-offs without incurring online LLM inference overhead. Experiments show that directly optimizing semantic relevance yields consistent offline gains in NDCG and positive online impact while preserving core engagement metrics. 

This work demonstrates that LLM supervision can be efficiently operationalized in scalable production search. Integrating semantic supervision into the ranking objective, rather than using LLMs solely as external judges or rerankers, offers a principled and scalable path toward more reliable and user-aligned e-commerce search.

Future work includes exploring richer semantic signals as inputs (e.g., query intent decomposition and attribute-level relevance), replacing global value model weights with query-adaptive weights, and extending the framework to incorporate long-term user satisfaction metrics.

\bibliographystyle{ACM-Reference-Format}
\balance
\bibliography{sample-base}










\end{document}